\documentclass[aps,pre,reprint,superscriptaddress,amsfonts,amssymb,amsmath,showpacs]{revtex4-1}

\usepackage{graphicx}
\usepackage{hyperref}
\usepackage{color}

\newcommand{\red}[1]{}
\newcommand{\blue}[1]{#1}

\newcommand{\del}[1]{}
\newcommand{\new}[1]{#1}

\providecommand{\e}[1]{\ensuremath{\times 10^{#1}}}

\begin{document}

\title{Anomalous capillary filling and wettability reversal in nanochannels}

\date{\today}

\author{Simon Gravelle}
\affiliation{Institut Lumi\`ere Mati\`ere, UMR5306 Universit\'e Lyon 1-CNRS, Universit\'e de Lyon 69622 Villeurbanne, France}
\author{Christophe Ybert}
\affiliation{Institut Lumi\`ere Mati\`ere, UMR5306 Universit\'e Lyon 1-CNRS, Universit\'e de Lyon 69622 Villeurbanne, France}
\author{Lyd\'eric Bocquet}
\affiliation{LPS, UMR CNRS 8550, Ecole Normale Sup\'erieure, 24 rue Lhomond, 75005 Paris, France}
\author{Laurent Joly}
\email{laurent.joly@univ-lyon1.fr}
\affiliation{Institut Lumi\`ere Mati\`ere, UMR5306 Universit\'e Lyon 1-CNRS, Universit\'e de Lyon 69622 Villeurbanne, France}

\begin{abstract} 
This work revisits capillary filling dynamics in the regime of nanometric to subnanometric channels\del{, a situation where a breakdown of the standard continuum framework is expected}. Using molecular dynamics simulations of water in carbon nanotubes, we show that for tube radii below one nanometer, both the filling velocity and the Jurin rise vary non-monotonically with the tube radius. Strikingly, with fixed chemical surface properties, this leads to confinement-induced reversal of the tube wettability from hydrophilic to hydrophobic for specific values of the radius. 
\del{We show that this effect originates in the non-monotonic behavior of the disjoining pressure associated with molecular fluid structuring.}
\new{By comparing with a model liquid metal, we show that these effects are not specific to water. Using complementary data from slit channels, we then show that they can be described using the disjoining pressure associated with the liquid structuring in confinement.}
This breakdown of the \new{standard} continuum framework is of main importance in the context of capillary effects in nanoporous media, with potential interests ranging from membrane selectivity to mechanical energy storage. 
\end{abstract}




\maketitle

\section{Introduction}

Capillary rise is a standard phenomenon of \del{continuum} fluid transport, coupling capillarity to hydrodynamics \cite{deGennes2013,Zhmud2000,Huber2015}. In view of its importance for many applications, this phenomenon has been thoroughly investigated for more than an century \cite{Lucas1918,Washburn1921,Bosanquet1923}.
Yet it still raises many questions, as a growing number of situations of interest now concerns capillary effects arising in nanometric to subnanometric channels \cite{Huber2015,Supple2003,Dimitrov2007,Caupin2008,Schebarchov2008,Gruener2009,Oyarzua2015,Vo2015,Vincent2015}. This is associated with the fast development of nanofluidics \cite{Bocquet2014} and the renewed activities around membranes and nanoporous media arising from water resources or energy 
problematics. At such scales new behaviors are expected \cite{Bocquet2014,Schoch2008, Sparreboom2009}, which jeopardize the standard treatment of capillary filling dynamics. This might be due to a breakdown of continuum hydrodynamics when approaching the liquid molecular scale \cite{Bocquet2010,Falk2015},  the onset of dominant slip effects \cite{Bocquet2007}, or  increasingly important deviations in interfacial properties \cite{Tolman1949,Bauer1999,Snoeijer2008,VanHonschoten2010,Hofmann2010,Guillemot2012} \new{arising from the molecular structuring of ultraconfined liquids \cite{Lichter2007}.} 

In order to unveil some specificities of molecular scale capillary effects, we revisit in this contribution the classic problem of capillary filling in nanometric to subnanometric channels. Using molecular dynamics (MD) simulations, we focus on carbon nanotubes (CNTs) as a model confining system and on water as a representative liquid. Overall, we observe significant deviations from \new{standard} continuum predictions for tubes with diameters as large as $3$\,nm. In particular, for specific radii the fluid is expelled from the pore, in contrast with \new{standard} predictions. These remarkable behaviors are attributed to liquid \new{structuring} effects, which are able to reverse the effective wettability of the channel for a given system chemistry. \new{Additional simulations with a model liquid metal show that the anomalous capillary filling and the wettability reversal are not specific to water, which behaves as a simple liquid in that particular situation. Complementing these results with slit-geometries simulations, we then show that these deviations from the standard continuum framework can be captured by the disjoining pressure, usually introduced to account for fluid structuring in the context of homogeneous films.} Finally we discuss the consequences of this work for applications involving nanoporous media.

\section{Standard continuum description}
\label{sec:theory}

\begin{figure}
\includegraphics[width=\linewidth]{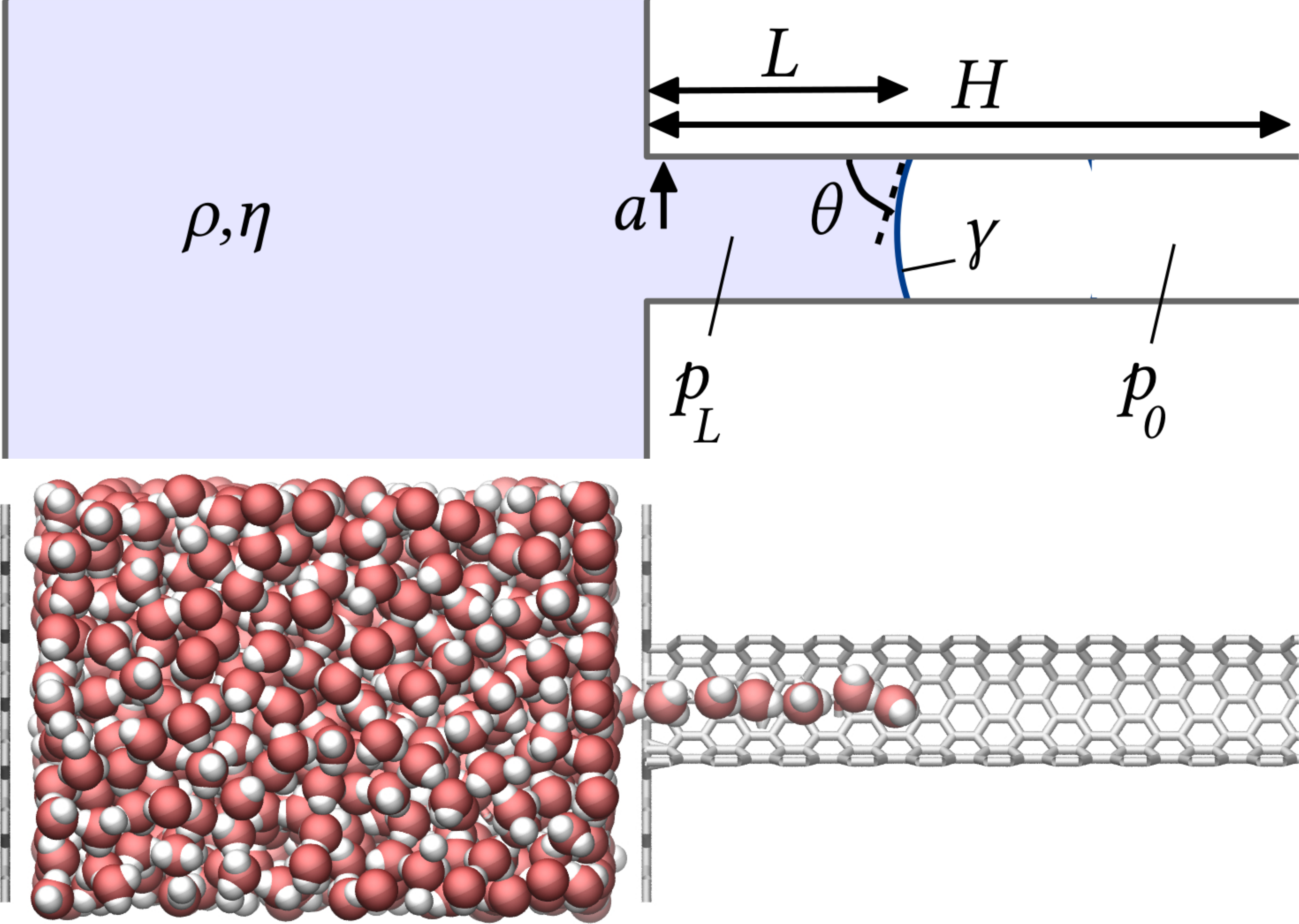}
\caption{Top: Schematic of the system. 
Bottom: Snapshot of a system used for the molecular dynamics (MD) simulations (tube radius $a_c = 3.9$\,\AA), made using VMD \cite{VMD}. The empty right reservoir is not shown.}
\label{fig:tube}
\end{figure}
We consider the capillary filling of a CNT (radius $a$, length $H$) with water (density $\rho$, viscosity $\eta$), see Fig. \ref{fig:tube}. 
The liquid surface tension is denoted $\gamma$, the contact angle of the liquid on the solid $\theta$, and the length of the liquid column inside the tube $L$. 
The driving force for capillary filling is the Laplace pressure jump through the liquid/vapor meniscus at the extremity of the liquid column, which writes: 
\begin{equation}
\Delta p^\text{men} = p_0 - p_L = \frac{2 \Delta \gamma}{a}, 
\label{eq:pres_lap}
\end{equation}
where $\Delta \gamma = \gamma \cos \theta$. 
%
%
The filling dynamics can then be limited by viscous friction inside the tube and at the tube entrance, or by inertia. 
At the macroscale, the tube inner friction generally dominates, leading to the standard Lucas-Washburn law \cite{Lucas1918,Washburn1921}. 
However, experimental and numerical work has shown that water flows with an ultralow friction in CNTs\new{, due to giant liquid/solid slip} \cite{Kannam2013}. This results in a plug-flow regime with a flat velocity profile inside the tubes, and a negligible inner viscous dissipation as compared to entrance dissipation \cite{Sisan2011}. 
Moreover, inertial effects can be neglected compared to viscous ones in the considered situation \cite{Joly2011}. Therefore the filling dynamics is only limited by entrance dissipation, 
inducing a pressure drop $\Delta p^\text{ent}$ related to the flow rate $Q$ according to Sampson equation:  
\begin{equation}
\Delta p^\text{ent} = \frac{C \eta}{2 a^3} \times Q,
\label{eq:pres_ent}
\end{equation}
%
with $C$ a numerical prefactor depending on the entrance geometry and hydrodynamic boundary condition \cite{Sampson1891,Gravelle2013}. 
Using that $Q=\pi a^2 v$ in the plug-flow regime, and writing that $\Delta p^\text{men}=\Delta p^\text{ent}$, one finds that the liquid fills the tube with a constant velocity $v_c$ (capillary velocity) given by: 
\begin{equation}
v_c = \frac{4 \Delta \gamma}{\pi C \eta}.
\label{eq:vc}
\end{equation}
\blue{At first sight, the capillary velocity seems independent of the tube radius. At the nanoscale however, the spherical shape of carbon atoms effectively chamfers the tube entrance, leading to a radius-dependent decrease of the Sampson coefficient $C$. This effect is well described by continuum hydrodynamics down to single-file flows \cite{Gravelle2014}, and $C(a_c)$ can be computed using finite element calculations, see the Appendix.} 

\new{Equation \eqref{eq:vc} accounts for specific nanoscale effects (liquid/solid slip and atomic chamfering of the tube entrance) but it ignores the discrete nature of the liquid and its possible structuring in confinement. We will therefore refer to Eq. \eqref{eq:vc} as the standard continuum prediction \footnote{Note however that structuring effects can be accounted for in continuum descriptions, \textit{e.g.} in the density functional theory framework  \cite{Bauer1999,Hofmann2010}.}.}

\section{Molecular dynamics simulations}

We now test the robustness of \new{the standard continuum prediction} at the molecular scale using MD simulations. 

\subsection{Systems}

We considered an initially empty CNT bounded by pierced graphene sheets, in contact with a water reservoir (left end) and an empty reservoir (right end); Fig. \ref{fig:tube} shows the water reservoir and the left tube entrance. 
The CNT length was 10\,nm, with radii $a_c$ varying between 3.9\,\AA\ and 24\,\AA.
Note that $a_c$ refers to the position of carbon atoms.
We showed in previous work that the effective radius $a$ seen by water molecules is smaller: $a_c -a \approx 2.5$\,\AA\ \cite{Gravelle2014}.
Periodic boundary conditions were imposed in all directions and a third graphene sheet (the left one in Fig. \ref{fig:tube}) was used as a piston. 
This piston prevented evaporation from one reservoir to the other
through periodic boundaries. We also used it to stop the filling process in order to measure the meniscus pressure jump in static conditions, see below. 

The simulations were performed with LAMMPS \cite{LAMMPS}.
We used the TIP4P/2005 model for water \cite{Abascal2005b} and the
AMBER96 force field for the carbon-oxygen interactions, i.e. a
Lennard-Jones potential with parameters $\epsilon_{CO} =
0.114$\,kcal/mol and $\sigma_{CO} = 0.328$\,nm \cite{Cornell1995}. 
Long-range Coulombic interactions were computed using the
particle-particle particle-mesh (PPPM) method. 
Water molecules were held rigid using the SHAKE algorithm. 
The equations of motion were solved using the velocity Verlet
algorithm with a timestep of $2$\,fs. 

The positions of the carbon atoms (wall+CNT) were fixed \footnote{Simulations with flexible and fixed walls were shown to give similar results for the statics and friction of confined liquids in previous work \cite{Alexiadis2008,Thomas2009,Werder2003}.}.
\new{Water molecules located more than 5\,\AA\ away from the membrane enclosing the tube entrance inside the left reservoir were kept at a temperature of 300\,K using a 
Berendsen thermostat \cite{Berendsen1984}.} 
The cut-off value $r_c$ for the Lennard-Jones potential was taken equal to 12\,\AA, 
which is slightly larger than the usual 10\,\AA\ \cite{Abascal2005b,Cornell1995}.
Indeed, the capillary velocity was found to depend on the value of $r_c$, 
converging only for $r_c \geq 12$\,\AA.
Moreover, the box size along the $x$ and $y$ directions was taken
sufficiently large to ensure that interactions between image CNTs did
not affect the velocity measurements, typically twice the tube
diameter \cite{Jensen2014}. 
Finally we made sure that the reservoir was larger than $10$ times
$a_\text{c}$ along the tube axis. Finite element calculations
indicated that, in that configuration, the error on the entrance
pressure drop due to finite size effects is lower than $0.25\,\%$
\cite{Gravelle2014}. 

\subsection{Data acquisition}

\begin{figure}
	\centering{
		\includegraphics[width=0.8\linewidth]{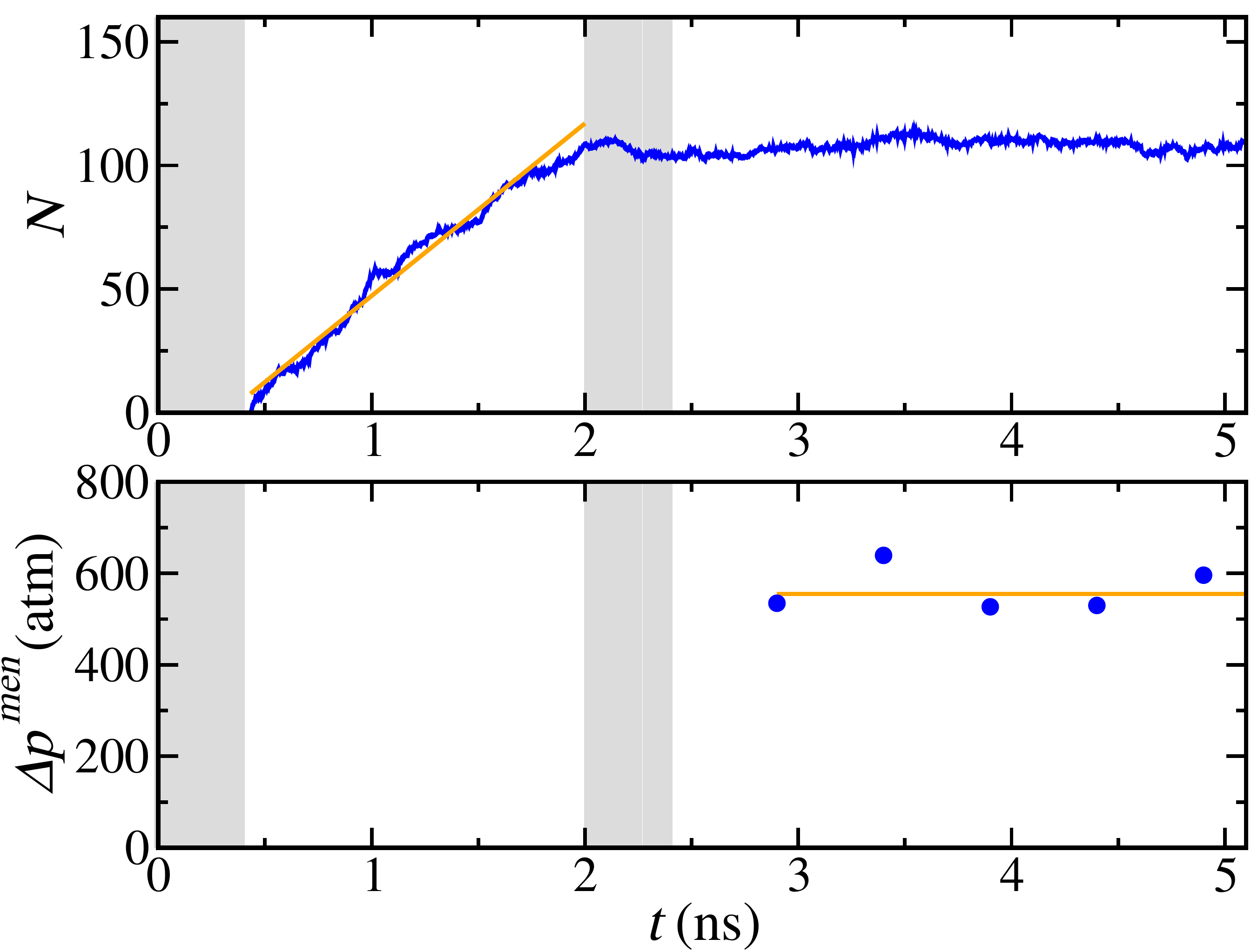}}
	\caption{Example of data acquisition (tube radius $a_c$ = 5.1 \AA). Up: number of molecules inside the tube $N$ as a function of time. 
		Down: measured meniscus pressure jump $p_0 - p_L$. 
		MD results are in blue, linear fits in orange and equilibration phases are gray areas. 
	}
	\label{fig:rempli}
\end{figure}
The protocol of data acquisition has been divided in four stages in
order to extract the two quantities of interest, which are the
capillary velocity (dynamic regime) and the meniscus pressure jump (static regime).
Before each measurement stage, equilibration stages were performed. 
Let us now describe those different stages as well as the
corresponding measured quantities. 
Water molecules are initially placed on a simple cubic lattice with equilibrium density. 
The system is equilibrated during 0.4\,ns, with a plug at the tube entrance to prevent water from entering. 
The plug is then removed and the evolution of the number of molecules inside the tube $N$ is recorded as a function of time.
This constitutes the dynamic regime of measurement. For a
given tube radius, $N$ increased linearly with time during the
simulation, corresponding to a constant filling velocity (capillary velocity), as expected from the standard continuum
prediction (see Sec. \ref{sec:theory}). 
The capillary velocity $v_c$ could thus be defined as
\begin{equation}
v_c = \frac{\mathrm{d} N}{\mathrm{d} t} \frac{1}{\lambda} , 
\end{equation}
where $\mathrm{d}N/\mathrm{d}t$ is the temporal derivative of the number of molecules inside the tube $N$ fitted during the dynamic stage and $\lambda$ is the linear density of fluid inside the tube, measured once the tube is filled. 
The dynamic regime lasts until the tube is partially filled (typically 1 or 2\,ns, depending on the tube radius).
At that moment, the piston is frozen, which prevents water molecules
to continue filling the tube. 
After another equilibration stage of 0.4\,ns, the pressure inside the liquid (measured by recording the total force on the piston, divided by its surface) is recorded during 2.5\,ns: this constitutes the static measurement regime. 
In the absence of flow, this piston pressure corresponds to the
pressure on the liquid side of the meniscus, and can be used to
compute the static meniscus pressure jump.
An example for a tube radius $a_c$ = 5.1 \AA\ is plotted in Fig. \ref{fig:rempli}. 
For each set of numerical parameters, the simulations have been performed 5 times with different initial conditions; the resulting velocity and pressure values have been averaged and standard deviations have been calculated.

\section{Results}

\begin{figure}
\includegraphics[width=\linewidth]{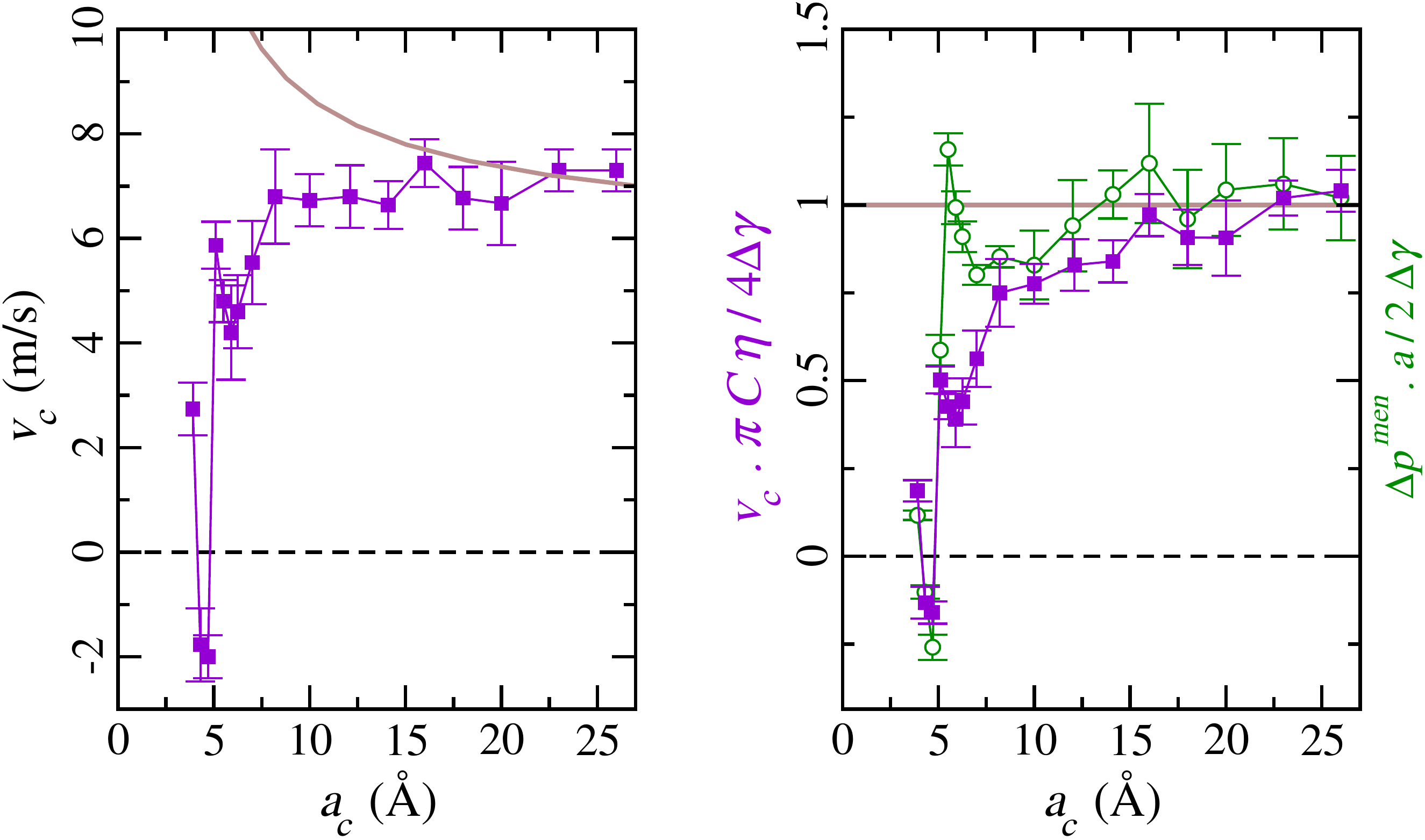}
\caption{Left: Capillary velocity $v_c$ as a function of the tube radius $a_c$, measured from MD simulations (violet squares). 
\blue{The brown line corresponds to the continuum prediction $v_c = 4 \Delta \gamma / \{\pi C(a_c) \eta \}$, with $C(a_c)$ computed using the finite element method (see text for details), $\Delta \gamma=14$\,mN/m and $\eta=0.855$\,mPa\,s.} 
Right: Normalized capillary velocity $v_c \cdot \pi C(a_c) \eta / (4 \Delta \gamma)$ (full violet squares) and meniscus pressure jump $\Delta p^\text{men} \cdot a / (2 \Delta \gamma)$ (open green circles), with $a$ the effective radius of the tube (see text). 
}
\label{fig:vitesse}
\end{figure}
Figure \ref{fig:vitesse} compares the measured capillary velocities $v_c$ with the \new{standard} continuum prediction, Eq. \eqref{eq:vc}, for various tube radii $a_c$. 
\blue{In the prediction, we included the variation of the Sampson coefficient $C$ with the tube radius $a_c$, computed in previous work using the finite element method \cite{Gravelle2014}, see the Appendix.}

\new{The standard} continuum prediction should hold for the largest tubes \cite{Joly2011,Giovambattista2015} and
for $a_c >15$\,\AA, MD results indeed agree with Eq. \eqref{eq:vc} when taking for the viscosity the tabulated value for TIP4P/2005 water and for $\Delta \gamma$ a value of 14\,mN/m. 
Below $15$\,\AA\ however, the results deviate from Eq. \eqref{eq:vc}, represented as the solid line in Fig. \ref{fig:vitesse}.
%
Such deviations are not unexpected, 
as interfacial properties are predicted to depart from their large-scale values at the nanoscale \cite{VanHonschoten2010}, as encompassed for instance in the Tolman length correction to the  surface tension \cite{Tolman1949} \new{or in line tension effects \cite{Getta1998} \footnote{Line tension should play no role in the cylindrical and slit geometries we considered, because a change in the contact angle does not modify the length of the contact line.}}. 
However 
two striking features appear in Fig. \ref{fig:vitesse} for radii $a_c<7.5\,$\AA. First the decrease in capillary velocity is \textit{not monotonous} with the tube radius, with local maxima of $v_c$ measured for $a_c$ around 5.1\,\AA\ and 3.9\,\AA. Second for radii $a_c$ = 4.3 and 4.7\,\AA, the capillary velocity $v_c$ becomes negative \footnote{In practice, no spontaneous filling is observed and the initial condition is modified by preparing the tube in the already filled configuration. From there, the measured evolution corresponds to a capillary emptying with the water being ejected from the nanotube} (reverse capillary flow).

\begin{figure}
\centering\includegraphics[width=\linewidth]{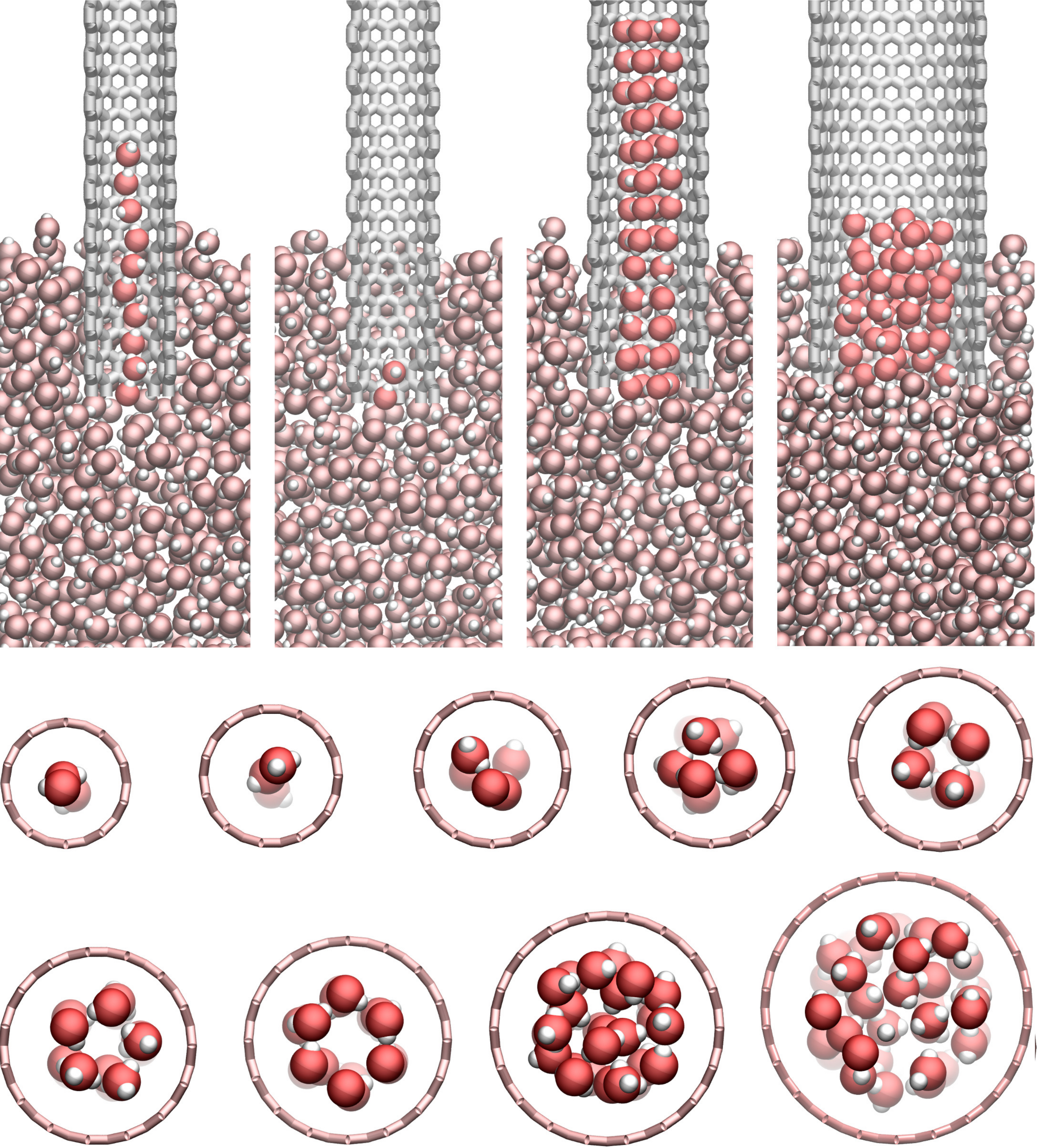}
    \caption{
Top: Ascension under a gravity field (Jurin-like numerical experiment) inside subnanometric nanotubes.
Tube radii from left to right: 3.9, 4.3, 5.5, and 8.2\,\AA. 
Bottom: Snapshot of water molecules inside CNTs of various radii, $a_c$ = 3.9, 4.3, 4.7, 5.1, 5.5, 5.9, 6.2, 7.0, and 8.2\,\AA.}
\label{fig:jurin_law}
\end{figure}

The observed anomalous capillary filling could be caused by deviations of both hydrodynamics and capillarity from their \new{standard} continuum behavior. 
However, \blue{as discussed above, the continuum description of entrance hydrodynamics remains valid down to the single file regime. Moreover,} hydrodynamics alone cannot explain the appearance of a reverse capillary flow, pointing to the crucial role of capillarity. 
%
%
To illustrate that scenario, we performed a Jurin-like numerical experiment, by comparing the final height of water within vertical nanotubes, under 
a gravity field $g$, 
Fig. \ref{fig:jurin_law}. For illustrative purpose, we chose an unphysically large value of $1.7 \e{13}$\,m\,s$^{-2}$ so that the 
capillary rise remained nanometric.
%
Jurin's law predicts that the filling height should scale as the inverse of the tube radius. Fig. \ref{fig:jurin_law} shows a radically different picture, confirming the dynamical measurements. In particular, a tiny change in radius (from 3.9 to 4.3\,\AA) can switch the pore behavior from hydrophilic to hydrophobic!

In order to assess quantitatively the role of capillarity in controlling the filling dynamics, we measured independently the meniscus pressure jump in static simulations, using a fixed piston to stop the meniscus in a half-filled nanotube position \footnote{For hydrophobic tubes, water is forced to enter the channel by applying a pressure to the piston, until the tube is half-filled.}. 
Figure \ref{fig:vitesse} presents MD results for the previous capillary velocity and for the present static meniscus pressure jump, both normalized by their \new{standard} continuum prediction, Eqs. \eqref{eq:vc} and \eqref{eq:pres_lap} respectively.  
The two characterizations agree quantitatively over almost the entire range of radii, showing identical features of local minima/maxima 
together with the wettability inversion. This confirms that molecular scale deviations from classical expectations arise due to a breakdown of the \new{standard} continuum prediction for the driving pressure term. 
Let us note that some discrepancy between the dynamic and static approaches can be seen in the 5.5 to 7\,\AA\ radius range. 
\red{Although the exact origin of these differences remains unclear, it is possible to identify at least two effects that would  impact differently dynamic and static problems. First, this difference could originate from entrance dissipation, 
which contributes to setting the capillary velocity (dynamic situation) but not the meniscus pressure jump (static situation). However, previous simulations \cite{Gravelle2014} have shown that continuum hydrodynamics describes entrance effects well
down to the single-file regime, see Fig. S4 in the SM.} 
As we will show in the following, we attribute the confinement-induced evolution of the capillary velocity to structuring effects in the liquid. 
The difference between static and dynamic cases could thus be associated with changes in the fluid structuring under both conditions.

\section{Discussion}

Overall, subnanometric tubes showed anomalous capillary imbibition departing from \new{standard} continuum expectations. These deviations, peaking at special radii ($a_c$ = 4.3-4.7\,\AA) for which a wettability reversal is observed, are controlled by changes in the driving pressure jump at the static meniscus. We now discuss the origin of this molecular-scale effect. 
We first note that other works suggest the existence of special radii for water in CNTs. For instance, water molecules must overcome a larger free energy barrier to enter a CNT with radius $a_c=4.6$\,\AA\ than to enter smaller and larger tubes \cite{Song2009, Corry2008}. Similarly, water can be stabilized in CNTs by very different mechanisms depending on the tube radius \cite{Pascal2011}.
%
To probe whether the observed capillary phenomenon was water-specific, we conducted complementary simulations of CNTs imbibition by a simpler liquid. We considered a model liquid metal, which lacks directional interactions and hydrogen-bonding effects. 
Compared to the even simpler Lennard-Jones liquids, liquid metals have the advantage of their large cohesion which allows the liquid to resist to the high negative pressures at play in (sub)nanometric capillary filling. 

\subsection{Capillary filling of CNTs by a liquid metal}

To complement the results for water, we
simulated the capillary filling of CNTs by a liquid metal. 
Again, the capillary velocity and the meniscus pressure jump have been
recorded for various tube radii ($a_c = 3.5$ to 23\,\AA). 
%
%
Liquid metals are commonly described by the embedded atom method (EAM) \cite{Daw1983}.
Here, we based our model liquid metal on the liquid gold EAM model developed by Grochola et al. \cite{Grochola2005}. 
The fluid temperature was set to 1200\,K and we adjusted the so-called 
distance space parameter $dr=0.64\,10^{-3}$\,\AA\ to reproduce water density, thus making comparisons easier.
We measured both the liquid metal viscosity, $\eta_m = 30 \pm
4$\,mPa\,s, and surface tension, $\gamma_m = 777 \pm 2$\,mN/m. 
Finally the metal and carbon atoms interacted through a Lennard-Jones
potential with an interaction parameter $\epsilon_{MC} = 0.052$\,kcal/mol, while keeping $\sigma_{MC}=\sigma_{OC}=3.28$\,\AA, where $M$ refers to metal, $C$ to carbon and $O$ to oxygen.

\begin{figure}
	\centering{
		\includegraphics[width=\linewidth]{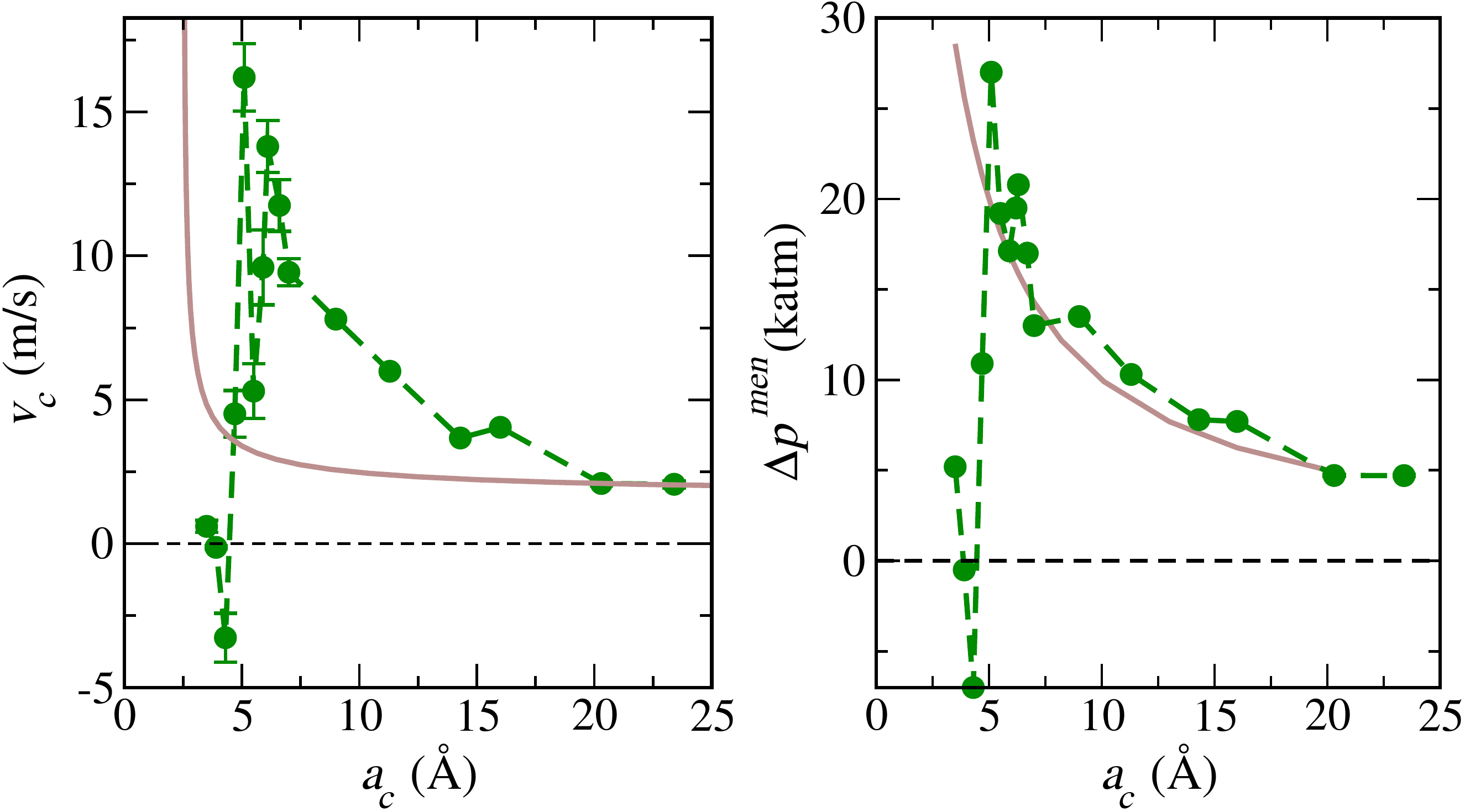}}
	\caption{Left: capillary velocity of the model liquid metal $v_c$ as
		a function of the CNT radius $a_c$.
		Right: meniscus pressure jump $p_0 - p_L$ estimated from static
		measurements. 
		Green symbols represent MD results, and brown lines correspond to \new{standard} continuum predictions (see text).
	}
	\label{fig:meta}
\end{figure}
The capillary velocity $v_c$ and the meniscus
pressure jump $p_0-p_L$, measured for various radii, are plotted in Fig. \ref{fig:meta}, and compared with \new{standard} continuum predictions, respectively $4 \Delta \gamma/\pi \eta C$ for the velocity and $2 \Delta \gamma / a$ for the pressure.
As for liquid water, both the capillary velocity and meniscus pressure
jump deviate from \new{standard} continuum predictions for a tube radius below 1\,nm.
For example, for both liquids, negative velocities are found for tubes with radii $a_c \sim 4.3 - 4.7$\AA, as well as positive velocities for $a_c$ = 3.9 \AA~ and $a_c  \ge$ 5.1 \AA.
Altogether, our results indicate that the wettability reversal and
non-monotonous evolution of the capillary velocity are not specific to water, and are obtained with simpler liquids too, irrespective for instance of the presence of directional constraints on the liquid structure. 
%
%
This global similarity despite the differences in detailed interaction properties suggests that a very generic mechanism is at play. This is the case of fluid structuring nearby solid surfaces that we consider below. 

\subsection{Structuring effects}

Figure \ref{fig:jurin_law} shows snapshots of water structuring inside CNTs. From single-file ($a_c=3.9$\,\AA) to quasi-continuum state ($a_c=8.2$\,\AA), the geometrical 
patterns shown by water molecules can be correlated to the velocity measurements presented in Fig. \ref{fig:vitesse}.
The three smallest tubes ($a_c=3.9$, $4.3$ and $4.7$\,\AA) can be filled by single-file water, but only the one with $a_c=3.9$\,\AA\ is hydrophilic, as seen in Fig. \ref{fig:vitesse}.
This probably arises from the increasing amount of vacuum from $a_c=3.9$ to $a_c =4.7$\,\AA, coming together with an energetically unfavorable increase of the water-gas contact surface.
For the larger radii $a_c=5.1$ and $5.5$\,\AA, side-by-side molecules are allowed, and those tubes show an increase in capillary velocity, consistently with the apparent small amount of vacuum present in these tubes. 
For tube radii $a_c=5.9$ and $6.2$\,\AA, a ring structure is observed, and the curve $v_c$ versus $a_c$ shows a local minimum, which could be related to the cost of having vacuum in the center of the tube. 
Finally, for radii $a_c=7.0$\,\AA\ and beyond, the fluid loses its structure, corresponding to a velocity that tends toward the \new{standard} continuum value ($\sim 7$\,m/s).

Formally, structuring effects can be described in terms of a disjoining pressure $\Pi_d$ that can be added to the \del{continuum} Laplace pressure $\Delta p^\text{Laplace}$  \cite{VanHonschoten2010}. The total pressure drop across the meniscus, previously defined as $\Delta p^\text{men} = p_0 - p_L$, writes now:
\begin{equation}
\Delta p^\text{men} = \Delta p^\text{Laplace}+ \Pi_d .
\label{eq:pl}
\end{equation}
$\Pi_d$ is a function of the ratio between the fluid molecule diameter $\sigma$ and the tube radius $a$.
While the expression of $\Pi_d$ as a function of the radius $a$ is not trivial for a cylinder, $\Pi_d$ can be expressed analytically for a fluid of hard spheres confined in a slit (2D) pore  as \cite{israelachvili2011intermolecular}: 
\begin{equation}
\Pi_d^\text{2D} (h) = - \rho_\infty k_B T \cos \left( 2 \pi h / \sigma \right) e^{- h / \sigma} , 
\label{eq:disj_pres}
\end{equation}
with $\sigma$ the fluid particle diameter, $\rho_\infty$ the fluid bulk density, and $h$ the 
distance between the walls. 
To test the validity of Eq. \eqref{eq:pl} together with Eq. \eqref{eq:disj_pres}, we simulated water filling slit nanochannels. 
The system was identical to the previous one, except that the channel was made of two planar graphene sheets, separated by a distance $h_c$ varied from 7.2 to 18\,\AA. As for CNTs, $h_c$ refers to the position of carbon atoms, and the effective height $h$ seen by water molecules is smaller: $h \approx h_c - 5$\,\AA.
The width $w$ of the system was larger than 10\,nm to avoid finite size effects in the lateral direction. 

\begin{figure}
\centering{
\includegraphics[width=\linewidth]{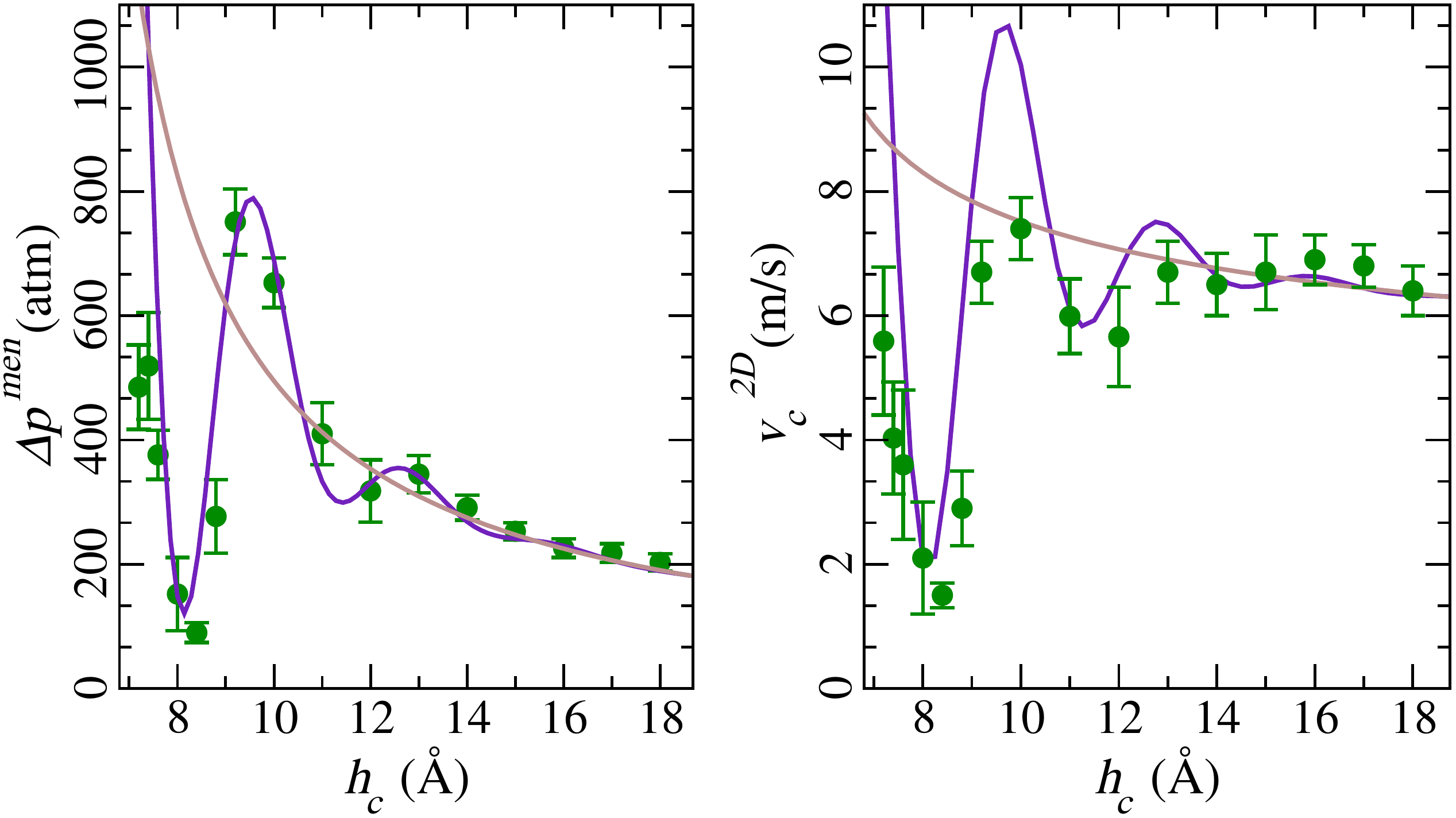}}
    \caption{
Capillary filling of slit graphitic channels. 
Symbols: total pressure drop $\Delta p^\text{men}$ measured in static simulations (left) and capillary velocity (right) as a function of the inter-plate distance $h_c$.
Brown line: continuum prediction (without structuring effects), using $\Delta \gamma = 25$\,mN/m. 
Violet line: extended prediction with structuring effects (see text for details). 
}
\label{fig:cap_2D}
\end{figure}
Figure \ref{fig:cap_2D} presents the results of static measurements of $\Delta p^\text{men}$, realized in the same manner as for CNTs. 
MD results can be fitted quantitatively using \blue{$\Delta p^\text{Laplace} = \Delta \gamma / h$ and} a modified expression of the disjoining pressure, with an added screening parameter in the exponential term and a larger value for $\rho_\infty$. These corrections account for the softer interactions between water molecules as compared to a hard sphere fluid \cite{israelachvili2011intermolecular}, and for the large contact density of water near a graphitic solid surface \cite{Tocci2014a}. 
Now turning to dynamics, \blue{the entrance pressure drop writes $\Delta p^\text{ent} = C^\text{2D} \eta v / (2 h)$, with the 2D Sampson coefficient $C^\text{2D}$ evaluated from finite element calculations in the spirit of Ref. \citenum{Gravelle2014}. The capillary velocity should then write 
$v_c^\text{2D} = (\Delta \gamma + h\,\Pi_d^{2D}) \times 2 / (C^\text{2D} \eta)$.} 
Figure \ref{fig:cap_2D}  shows a comparison between this prediction and MD simulations. 
As expected, the capillary velocity oscillates when varying the inter-plate distance $h_c$, and the qualitative agreement between theory and simulations is good. 
Note a quantitative difference for $h_c \sim$ 9-10\,\AA, 
reminiscent of a similar deviation observed in the case of the cylindrical geometry, see Fig. \ref{fig:vitesse}. 

\section{Conclusion}

To conclude, 
we have shown that in contrast with \new{standard} continuum expectations, the capillary velocity depends non-monotonically on the channel dimensions, specifically for tube radii below one nanometer, with the possibility of effectively reversing the wettability for special confinements. 
\new{We then have shown that these effects are associated with the structuring of water inside the pore.
Surprisingly, the anomalous capillary filling and wettability reversal are not specific to water -- which behaves as a simple liquid for this particular problem, as confirmed by results obtained with a model liquid metal. 
The observed phenomena 
can therefore be described with generic continuum models accounting for fluid structure at the molecular scale such as density functional theory} \cite{Bauer1999,Snoeijer2008,Hofmann2010}. 
\new{We have even shown that the deviations from the standard continuum framework could be described by the disjoining pressure, a continuum concept usually introduced in the context of homogeneous films.} 
On this basis, it is accordingly possible
to theoretically predict the capillary velocity down to the smallest confinement.
Altogether this shows that the \new{standard} continuum framework for capillary imbibition cannot be used {\it per se} to account for the capillary driven flows in subnanometric channels\new{, but that it can be refined with simple continuum tools accounting for liquid structuring}.

The present results are of importance in the study of the filling of nanoporous media.
Indeed, according to these results the hydrodynamic permeability of the medium should vary non-monotonically with the pore size distribution and the effect can be quite strong, particularly in the case of water filling nanoporous media with pore radii distributed around 3.9-5.1\,\AA.
A similar effect was actually reported experimentally for the capillary imbibition of water in a carbon nanotube \cite{Qin2011}, without an explanation up to now. 
Furthermore, the imbibition of nanoporous media should be disturbed by the present effect, leading to an effective porosity lower than the expected one. 
Finally, the present results could help to explain the selectivity of membranes,
with potential consequences for example in the understanding (and fabrication) of selective ion channels \cite{Gouaux2005}. 

\appendix*

\section{Entrance viscous dissipation}

\begin{figure}
	\vspace{9mm}
	\centering{\includegraphics[width=\linewidth]{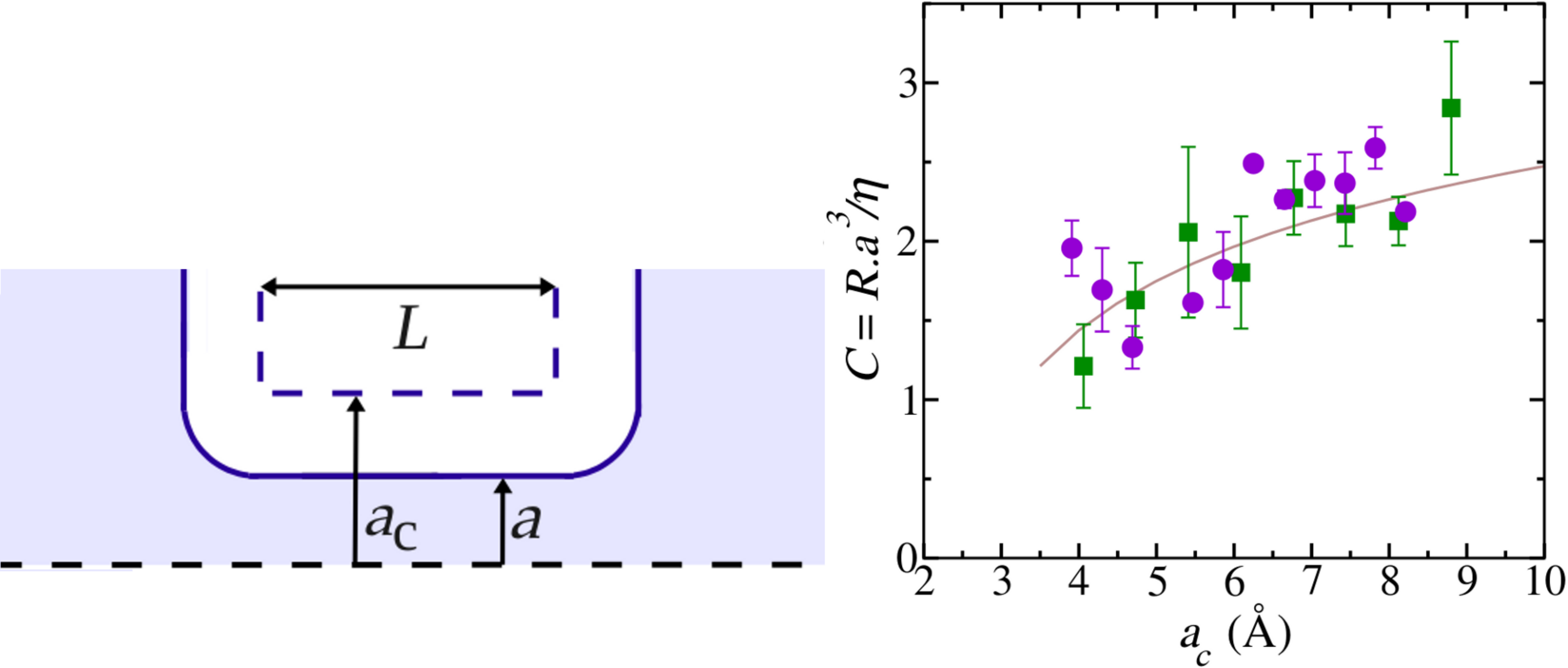}}
	\caption{\blue{Left: Geometry used for the finite element calculations. Two reservoirs are separated by a membrane containing a channel of radius $a$ and length $L$.
			The black dashed line corresponds to an axisymmetric condition.
			The full line represents the effective liquid/solid interface, where a
			perfect slip boundary condition applies. Note that the atomic chamfering of
			the entrance due to the spherical shape of carbon atoms is accounted
			for in the geometry used for the finite element calculations. 
			Right:} value of the $C$ coefficient calculated from the normalized
		hydrodynamic resistance, data extracted from
		Ref. \citenum{Gravelle2014}. 
		Circles represent MD results with armchair tubes, squares represent MD
		results with zigzag tubes, and the line is the hydrodynamic prediction
		computed with finite element calculations. 
	}
	\label{fig:Ccoef}
\end{figure}

From Ref. \citenum{Gravelle2014}, the viscous dissipation for a
liquid flowing through a carbon nanotube, including inner and entrance contributions, can be
obtained. The overall dissipation is dominated by entrance effects,
and entirely determined by the value of the numerical prefactor $C$,
see Eq.~\eqref{eq:pres_ent}. 
Values of $C$ for different tube radii are gathered in Fig. \ref{fig:Ccoef}.
\blue{First, note that the $C$ coefficient varies with the tube radius $a_c$.
	This variation comes from the chamfering of the entrance induced by
	the finite radius of carbon atoms, see Ref. \citenum{Gravelle2014}. 
	Moreover, note that continuum predictions are in good agreement with
	molecular dynamics results 
	for the description of the viscous entrance dissipation.
	Therefore,} viscous entrance effects cannot be the main reason of the deviations to the standard continuum predictions observed in the present paper.
\blue{In particular, viscous entrance effects cannot explain the
	reversal of the tube wettability.} 
Note however that these results have been obtained with a different water model (TIP3P).

\begin{acknowledgments}
S.G thanks Catherine Sempere for pointing out Ref. \citenum{Song2009}.
This research was supported by the European Research Council
programs Micromegas project.
\end{acknowledgments}

\end{document}